\documentclass[
    ,final
  ,numberedheadings 
  ,cmfonts          
  ,float
  ,subfigure
  ]
  {aipproc}

\usepackage{amsmath}
\usepackage{subfigure}
\layoutstyle{6x9}

\newcommand{\cf}[1]{{Fig.~\ref{#1}}}

\newcommand{\gmu}{\gamma^\mu}

\newcommand{\ie}{{\it i.e.}}

\renewcommand{\thefootnote}{\fnsymbol{footnote}}

\begin{document}
\title{Off-shell and non-static contributions to
heavy-quarkonium production}

\author{J.P.~Lansberg}{
address={Centre de Physique Th\'eorique, \'Ecole polytechnique, CNRS,  \\
91128 Palaiseau, France \\ and \\
Physique th\'eorique fondamentale, D\'epartement de  Physique, Universit\'e de  Li\`ege, \\ 
all\'ee du 6 Ao\^{u}t 17, b\^{a}t. B5, B-4000 Li\`ege~1, Belgium\\
E-mail: Jean-Philippe.Lansberg@cpht.polytechnique.fr}
}

\begin{abstract}
We have shown that if one relaxes the constraint that the quarks in a heavy quarkonium are at rest and
on-shell, new contributions to the discontinuity of the production amplitude appear. These can
be seen as a $s$-cut in the amplitude and are on the same footage as the classical cut
of the Colour-Singlet Model (CSM), where the heavy quarks forming the quarkonium are put
on-shell by hypothesis. We treat this cut in a gauge-invariant manner 
by introducing necessary new 4-point vertices, suggestive of the colour-octet
mechanism. We have further shown that this cut contributes at least as much as the LO CSM at large
$P_T$. However, the 4-point vertices cannot be totally constrained and an ambiguity remains 
to what concerns their actual contribution. Theoretical insights from meson photoproduction  
are discussed in that context.
\end{abstract}

\maketitle


\footnotetext{Presented at the 7th International Conference on Quark Confinement and Hadron Spectrum
(QCHS7), September 2-7 2006, Ponta Delgada, Portugal and at the 4th Meeting of the Quarkonium Working Group, June 27-30 2006, BNL, USA.}
\renewcommand{\thefootnote}{\arabic{footnote}}


More than ten years ago, the CDF collaboration~\cite{CDF7997a,CDF7997b}
brought to light the ``$\psi'$ anomaly'', \ie~an excessively large experimental cross section 
for $\psi'$ (and $J/\psi$) production
compared with theoretical expectations. No totally conclusive solution to this problem has been proposed so 
far (for recent reviews see~\cite{Brambilla:2004wf,Lansberg:2006dh}). Even though the 
Colour-Octet 
Mechanism (COM), coming from the application of NRQCD to heavy quarkonium, looked as
a very promising solution, it appears clearly that as long as fragmentation is the dominant
production contribution and the velocity-scaling rules of NRQCD hold, it cannot accommodate 
the polarisation measurements of CDF~\cite{Affolder:2000nn}, which show a non-polarised, if not 
slightly longitudinal, production.

In that context, we have felt the necessity to reconsider the appropriateness of 
the static and on-shell approximation of the Colour-Singlet Model (CSM)~\cite{CSM_hadron}.

In order to study properly non-static and off-shell effects, we have used a vertex function
as an input for the bound-state characteristics, whereas the Schr\"odinger wave function at the origin 
is used in the CSM and Long Distance Matrix Elements (LDME) of NRQCD enter the COM. We emphasise again
that we probe all the internal phase space of the quarkonium, and thus need a function, where these two
approaches simply need a constant factor.

In the case of $^3S_1$ quarkonium (noted $\cal Q$) production in 
high-energy hadronic collisions, we are to consider gluon fusion $gg\to {\cal Q} g$. 
Using the Landau equations~\cite{Landau:1959fi}, we have shown in~\cite{Lansberg:2005pc} that 
there are two families of contributions (see~\cf{fig:diag_LO_QCD} (a) and (b)): 
the first is the usual colour-singlet mechanism, where in the context of our model, 
we use a 3-point function $\Gamma^{(3)}_{\mu}(p,P) = \Gamma(p,P) \gamma_\mu$
 at the $Q\bar Q \cal Q$ vertex; the second family was never considered before. 
To simplify the study, we set $m>M/2$ so that the first cut does not contribute.
The functional form of $\Gamma(p,P)$ (gaussian or dipole) and its parameters
 have been discussed in details in~\cite{Lansberg:2005ed}.

\begin{figure*}
\centerline{\mbox{\includegraphics[height=4cm]{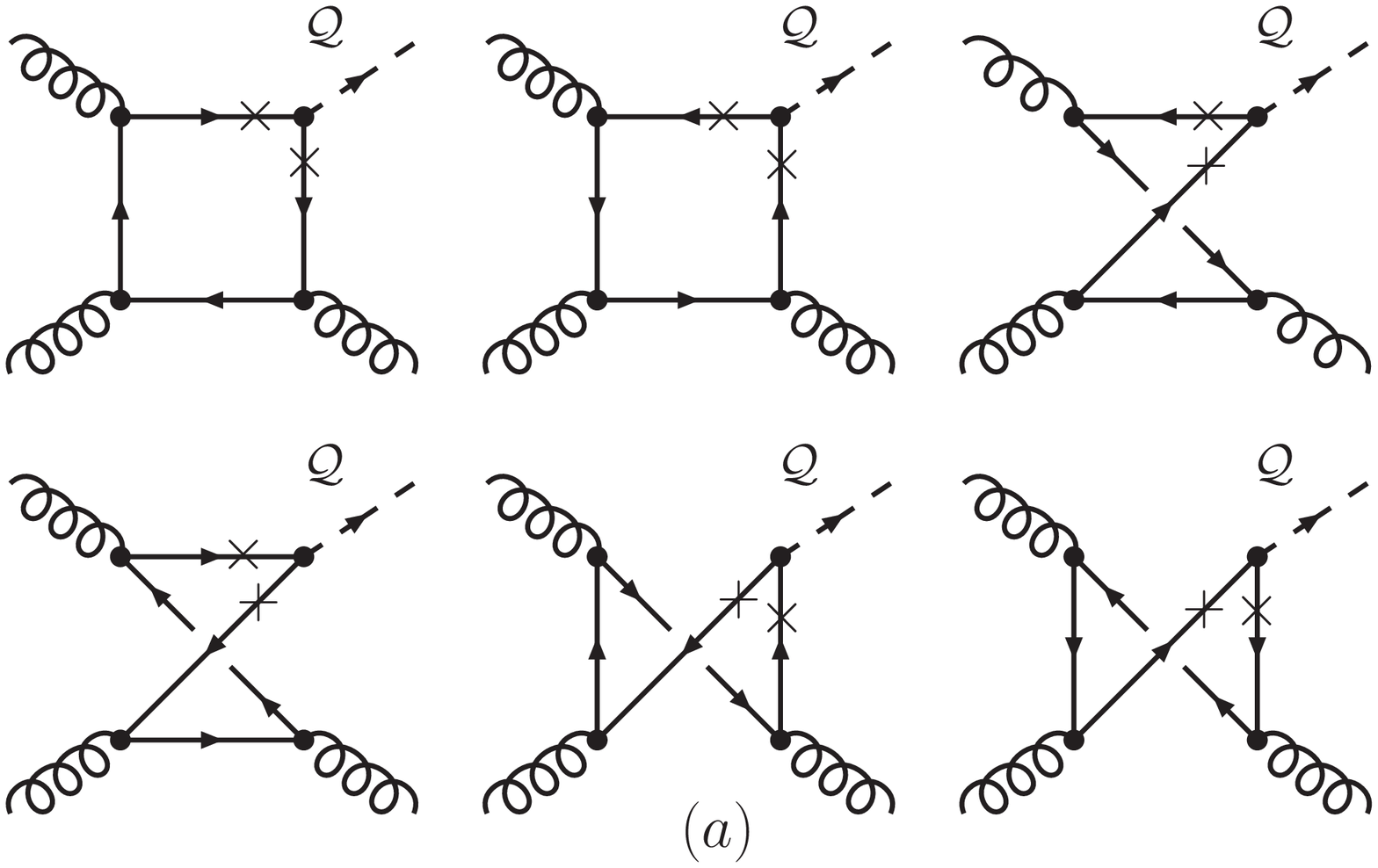}\quad
\includegraphics[height=4cm]{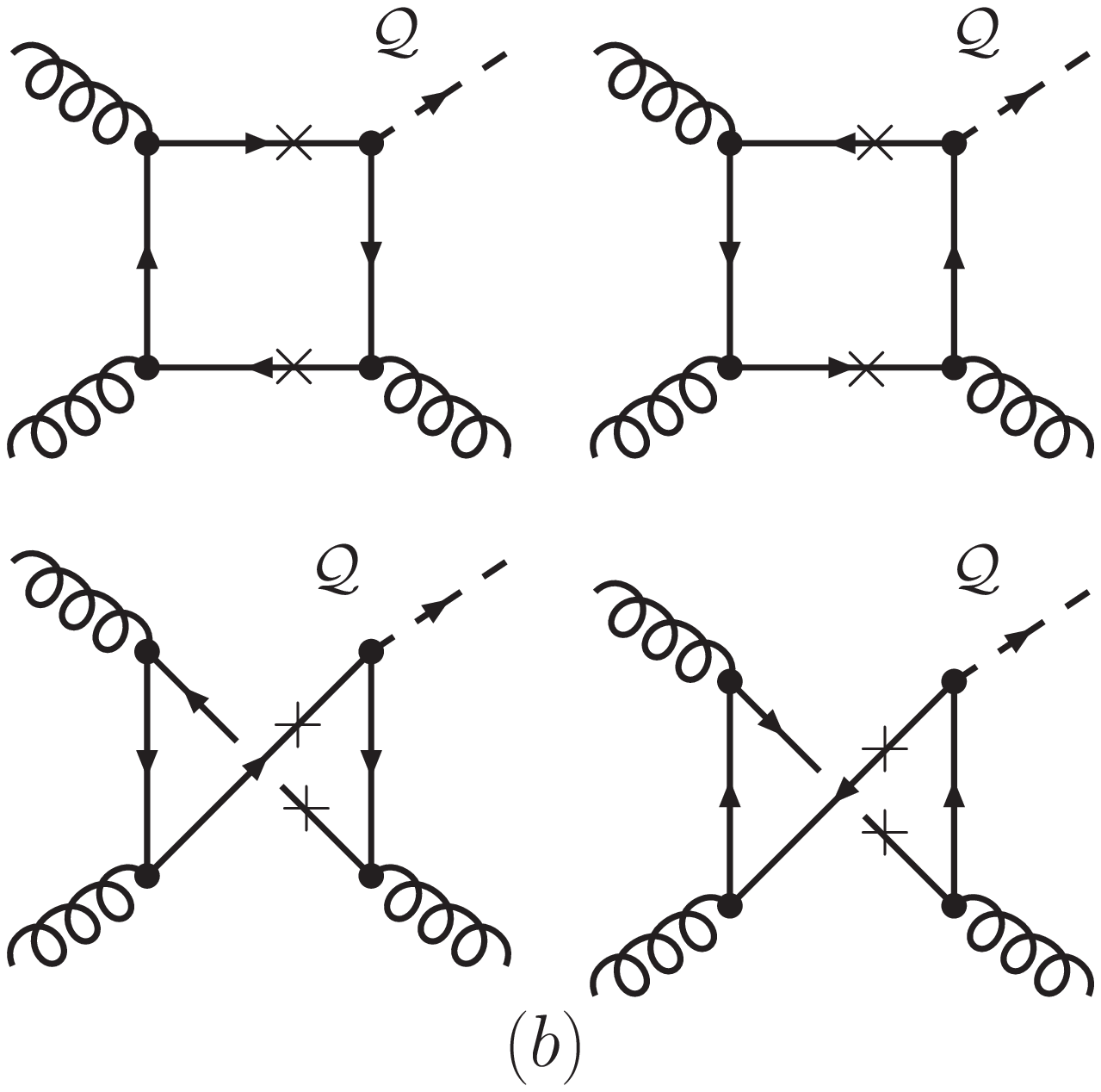}
\quad\includegraphics[width=3cm]{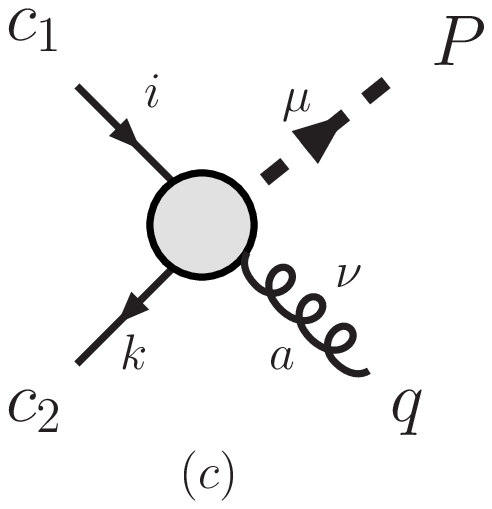}}}
\caption{The first family (a) has 6 diagrams and  the second family (b)  
4 diagrams contributing the discontinuity of $gg \to \!\!\ ^3S_1 g$ at LO in QCD.(c): the gauge-invariance restoring vertex, $\Gamma^{(4)}$.}
\label{fig:diag_LO_QCD}
\end{figure*}

In addition to the second family, one is driven -- to preserve gauge invariance (GI) -- to introduce
new contributions arising from the presence of 4-point vertices. Beside restoring
GI, these vertices have to satisfy specific constraints~\cite{Lansberg:2005pc,these,Drell:1971vx}. 
For the following 
simple choice for $\Gamma^{(4)}_{\mu\nu}(c_1,c_2,P,q)$
\begin{eqnarray}
-i g_s T^a_{ki} \left( \Gamma_1
-\Gamma_2\right)\left[{c_1^{\nu}\over (c_1-q)^2-m^2}
+{c_2^{\nu}\over (c_2+q)^2-m^2}\right]\gamma^\mu,
\end{eqnarray}
where $\Gamma_1\equiv \Gamma(2c_1-P,P)$, $\Gamma_2\equiv\Gamma(2c_2-P,P)$, the momenta and indices are as in~\cf{fig:diag_LO_QCD} (c), the results obtained for $J/\psi$ and $\psi'$ 
production at the Tevatron are exposed in
\cite{Lansberg:2005pc}. For the $J/\psi$ , we saw that the $s$-channel cut contributes at least as much
as the classical cut of the CSM at large $P_T$. 
In the $\psi'$ case, we employed the ambiguity upon the vertex-function 
normalisation~\cite{Lansberg:2005ed}   due to the node position to 
show that agreement with the data at low $P_T$ was conceivable. We further noticed that 
the $P_T$ slope was only slightly different from that of the data. 
This is at variance with what is widely believed since COM fragmentation 
(with a typical $1/P_T^4$ behaviour) processes are in agreement with experimental measurements.

Another possible Ansatz for $\Gamma^{(4)}_{\mu\nu}(c_1,c_2,P,q)$~\cite{haberzettl}, inspired from studies
of meson photoproduction~\cite{Haberzettl:1997jg,Haberzettl:1998eq,Davidson:2001rk} and which possesses a better behaviour at low momenta, is
\begin{eqnarray}
-i g_s T^a_{ki}\left(\frac{(2c_1-q)^\nu }{(c_1-q)^2-m^2}(\Gamma_2-F)+\frac{(2 c_2+q)^\nu }{(c_2+q)^2-m^2}(\Gamma_1-F)\right)\gmu
\end{eqnarray}
with $F=\Gamma_0-h (\Gamma_0-\Gamma_1) (\Gamma_0-\Gamma_2)$ ($\Gamma_0$ is the value of the vertex function
when $(c_1-q)^2=(c_2+q)^2=m^2$) and $h$ an arbitrary crossing-symmetric
function of the momenta. This will be studied in a future work. Applications to $\eta_c$ and 
$\eta'_c$ decays could  also be relevant in view of the possible anomaly of
$\eta'_c\to \gamma \gamma$ decay~\cite{Lansberg:2006dw}.

However, there exist further other choices for the GI restoring vertex. Interesting results 
are indeed obtained by studying the effects of autonomous vertices, which link
different suitable choices: they are GI alone and a priori unconstrained in normalisation. 
The latter can be fitted to described data from~\cite{CDF7997a,CDF7997b,Acosta:2001gv,Adler:2003qs} 
as shown in~\cite{these,Lansberg:2005gs}.

In conclusion, we have shown that it is possible to go beyond the on-shell and static approximations 
of the CSM. It 
may also be possible to extend the COM in the same manner. This necessitates the 
introduction of 4-point vertices due to the non-local 3-point vertex relevant for the non-static and
off-shell contributions.

By going deeper in the analysis, we have seen that the form of these 4-point vertices is not absolutely 
constrained, even after imposing necessary conditions to conserve crossing symmetry and the analytic 
structure of the amplitude. By exploiting this lack of constraint, we
have been able to reproduce the direct-production cross section for the $J/\psi$, 
$\psi'$ and $\Upsilon(1S)$ 
as measured at the Tevatron by CDF (and also at RHIC by PHENIX for $J/\psi$).



\bibliographystyle{aipprocl} 

\begin{thebibliography}{99}

\bibitem{CDF7997a}
F.~Abe {\it et al.}  [CDF Collaboration],
Phys.\ Rev.\ Lett.\  {\bf 79} (1997) 572.


\bibitem{CDF7997b}
F.~Abe {\it et al.}  [CDF Collaboration],
Phys.\ Rev.\ Lett.\  {\bf 79} (1997) 578.

\bibitem{Brambilla:2004wf}
N.~Brambilla {\it et al.}, {\it Heavy quarkonium physics}, CERN Yellow Report, CERN-2005-005, 
2005  Geneva : CERN, 487 pp 
[arXiv:hep-ph/0412158].

\bibitem{Lansberg:2006dh}
  J.~P.~Lansberg, {\it $J/\psi$, $\psi'$ and $\Upsilon$ production at hadron colliders: A review},
  Int.\ J.\ Mod.\ Phys.\ A {\bf 21} (2006) 3857
  [arXiv:hep-ph/0602091].


\bibitem{Affolder:2000nn}
T.~Affolder {\it et al.}  [CDF Collaboration],
Phys.\ Rev.\ Lett.\  {\bf 85} (2000) 2886
[arXiv:hep-ex/0004027].

\bibitem{CSM_hadron} 
C-H. Chang, 
Nucl. Phys.  {\bf B 172} (1980) 425;
R. Baier and R. R\"uckl, 
Phys. Lett.  B {\bf 102} (1981) 364;
R. Baier and R. R\"uckl,
Z. Phys.  {\bf C 19} (1983) 251.

\bibitem{Landau:1959fi}
  L.~D.~Landau,
  Nucl.\ Phys.\  {\bf 13} (1959) 181.


\bibitem{Lansberg:2005pc}
  J.~P.~Lansberg, J.~R.~Cudell and Yu.~L.~Kalinovsky,
  Phys.\ Lett.\ B {\bf 633}, 301 (2006)
  [arXiv:hep-ph/0507060].




\bibitem{Lansberg:2005ed}
  J.~P.~Lansberg,
  AIP Conf.\ Proc.\  {\bf 775} (2005) 11
  [arXiv:hep-ph/0507184].


\bibitem{these} J.~P.~Lansberg, {\it Quarkonium Production at High-Energy Hadron Colliders}, Ph.D. 
Thesis, ULg, Li\`ege, Belgium, 2005.


\bibitem{Drell:1971vx}
  S.~D.~Drell and T.~D.~Lee,
  Phys.\ Rev.\ D {\bf 5} (1972) 1738.

\bibitem{haberzettl}
H.~Haberzettl, private communication.


\bibitem{Haberzettl:1997jg}
  H.~Haberzettl,
  Phys.\ Rev.\ C {\bf 56} (1997) 2041
  [arXiv:nucl-th/9704057].

\bibitem{Haberzettl:1998eq}
  H.~Haberzettl {\it et al.},
  Phys.\ Rev.\ C {\bf 58} (1998) 40
  [arXiv:nucl-th/9804051].

\bibitem{Davidson:2001rk}
  R.~M.~Davidson and R.~Workman,
  Phys.\ Rev.\ C {\bf 63} (2001) 025210
  [arXiv:nucl-th/0101066].

\bibitem{Lansberg:2006dw}
  J.~P.~Lansberg and T.~N.~Pham,
  Phys.\ Rev.\ D {\bf 74} (2006) 034001
  [arXiv:hep-ph/0603113].

\bibitem{Acosta:2001gv}
D.~Acosta {\it et al.}  [CDF Collaboration],
Phys.\ Rev.\ Lett.\  {\bf 88} (2002) 161802.

\bibitem{Adler:2003qs}
S.~S.~Adler {\it et al.}  [PHENIX Collaboration],
Phys.\ Rev.\ Lett.\  {\bf 92} (2004) 051802
[arXiv:hep-ex/0307019].

\bibitem{Lansberg:2005gs}
  J.~P.~Lansberg,
  AIP Conf.\ Proc.\  {\bf 792} (2005) 823
  [arXiv:hep-ph/0507118].

\end{thebibliography}

\end{document}